\documentclass{emulateapj}

\usepackage{graphicx}
\usepackage{epsfig}
\usepackage{times}
\usepackage[]{natbib}
\usepackage{url}
\usepackage{color}
\usepackage{amssymb,amsmath}

\shorttitle{Eruptive microflares in the Sun.}
\shortauthors{Archontis and Hansteen}

\begin{document}

\title{Clusters of small eruptive flares produced by magnetic reconnection in the Sun}



\author{V. Archontis\altaffilmark{1}}
\author{V. Hansteen\altaffilmark{2}}


\altaffiltext{1}{School of Mathematics and Statistics, St. Andrews University, St. Andrews, KY169SS, UK}
\altaffiltext{2}{institute of Theoretical Astrophysics, University of Oslo, Norway}


\begin{abstract}

We report on the formation of small solar flares produced by patchy magnetic reconnection between interacting magnetic loops. 
A three-dimensional (3D) magnetohydrodynamic (MHD) numerical experiment was performed, where a uniform magnetic flux sheet was injected into a 
fully developed convective layer. The gradual emergence of the field into the solar atmosphere results in a network of magnetic loops, which interact dynamically 
forming current layers at their interfaces. The formation and 
ejection of plasmoids out of the current layers leads to patchy reconnection and the spontaneous formation 
of several small (size $\approx 1-2\,\mathrm{Mm}$) flares. 
We find that these flares are short-lived ($30\,\mathrm{s}-3\,\mathrm{min}$) bursts of energy in the range $O(10^{25}-10^{27})\,\mathrm{ergs}$, which is 
basically the nanoflare-microflare range. Their persistent formation and co-operative action and evolution leads to recurrent emission of 
fast EUV/X-ray jets and considerable plasma heating in the active corona.

\end{abstract}



\section{Introduction}

Observations \citep[e.g.][]{lin84,porter87,hannah08} have revealed the existence of numerous microflares 
(transient brightenings with energy $O(10^{27})\,\mathrm{ergs}$ and size smaller than the standard  flares \citep[e.g.][]{priest02,shibata11} on the Sun. 
The areas around microflares are often bright in X-rays, which implies plasma heating \citep[e.g.][]{porter87,hudson91}. 
Thus, microflares have been considered as possible sources for heating the solar corona \citep[e.g.][]{lin84,porter87}, subject to their 
occurence rate and energy release. 
On theoretical grounds, \citet[]{parker88} suggested that the active X-ray corona consists of numerous nanoflares ($O(10^{24})\,\mathrm{ergs}$, largest 
nanoflares approaching $10^{26}-10^{27}\,\mathrm{ergs})$ 
and that microflares could be made up of several nanoflares.
Also, previous two-dimensional simulations \citep[]{yoko95} and observations \citep[]{chae99} have shown that X-ray jets and 
cooler $H_\alpha$ surges can be emitted from microflares, possibly due to reconnection between emerging and pre-existing 
coronal magnetic fields. All the above suggest that although nano/micro-flares are small-scale events, they have a great influence on the 
solar atmosphere. However, their 3D formation, evolution and energetics are not well known. 
Moreover, the existing standard flare models \citep[e.g.][]{sweet69,kopp76,shibata95} elaborate the onset of larger and single flares in 
coherent (almost ‘monolithic’) current sheets and, therefore, cannot address simultaneously the onset of nano/micro-flares and the intermittent 
heating \citep[e.g.][]{machado88,parker88,isobe05,nishizuka10} of the solar corona. 

Here we report radiative MHD
simulations, showing that small flares are formed naturally 
by patchy reconnection, which is triggered by the eruption of plasmoids \citep[e.g.][]{tsuneta97,shibata01,arc06} in fragmented current sheets,
between interacting magnetic bipoles \citep[e.g.][]{machado88,hanaoka96}.
The fragmentation of currents explains naturally the ubiquitous 
intermittent heating and filamentary nature \citep[e.g.][]{howard72,isobe05} of the solar corona. Moreover, we find that the frequent onset and 
co-operative action of small flares dump enough energy in the solar atmosphere, sufficient to accelerate and heat the plasma in the active corona.

\section{The model}
\begin{figure*}[t]
\centering
\includegraphics[width=0.78\linewidth]{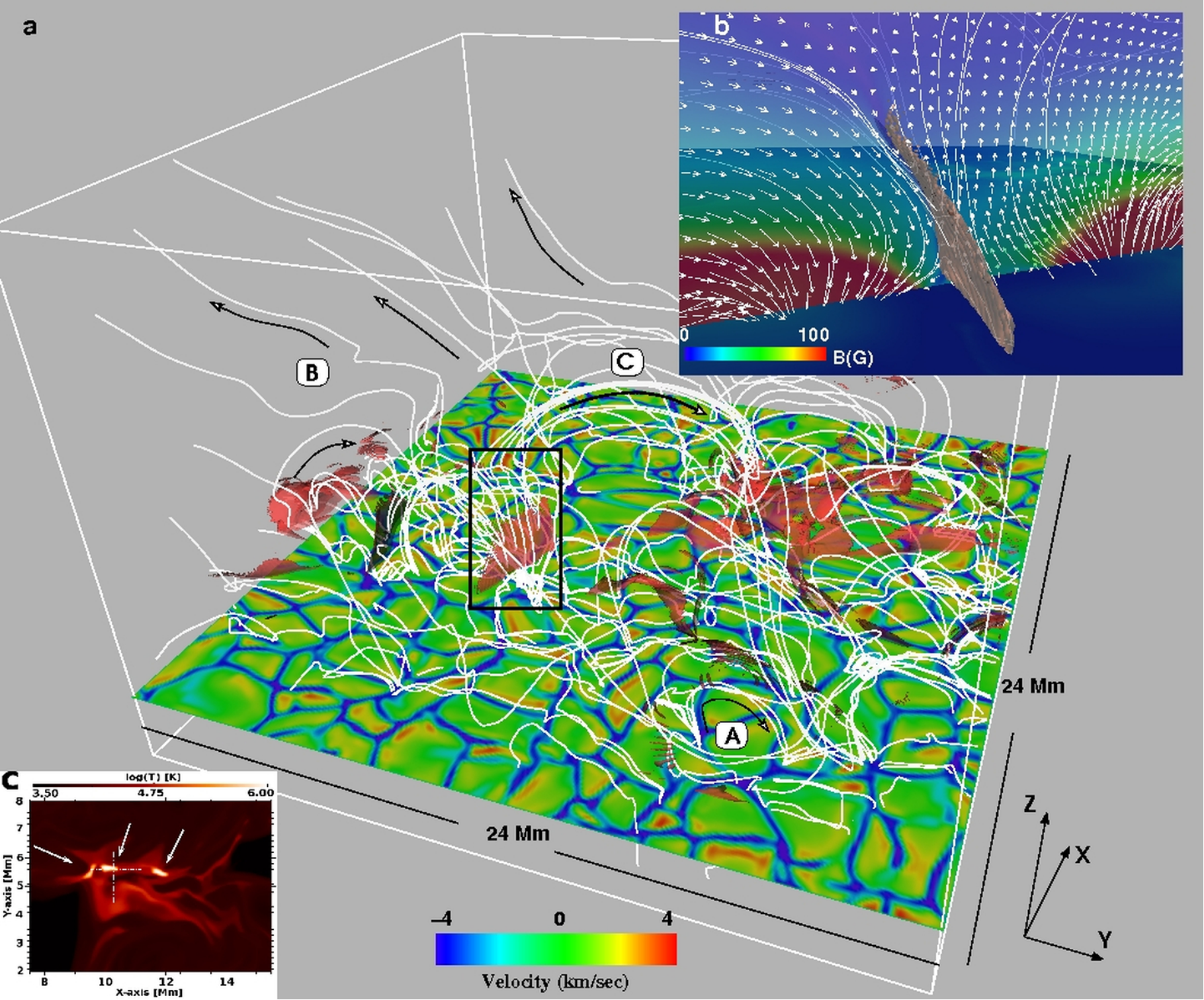}\caption{
{\bf a)} Fieldlines show the magnetic topology of the emerging field at $t = 7800\,\mathrm{s}$. 
Current sheets (transparent isosurfaces with current density ($0.042\,\mathrm{A}\,\mathrm{m}^{-2}$) are formed between interacting loops. 
Highly bent fieldlines (e.g. B) is the result of reconnection between emerging and pre-existing magnetic fields. 
The arrows (black) show the direction of the magnetic field along the fieldlines. 
The horizontal slice ($z = 0.28\,\mathrm{Mm}$) shows the photospheric flows as a result of granular convection. 
{\bf b)} Close-up of the interface current, which is outlined by the rectangular inset in a). 
The fieldlines of the neighbouring magnetic loops start to reconnect at the current sheet. 
The full magnetic field vector (arrows in the range $3.4 \le y \le 7.3\,\mathrm{Mm}$, $0.62 \le z \le 4.28\,\mathrm{Mm}$, $x = 10.35\,\mathrm{Mm}$) 
shows the oppositely directed and highly sheared field across the current. 
{\bf c)} Temperature distribution at $z = 4.32\,\mathrm{Mm}$ ($t = 8770\,\mathrm{s}$) reveals a cluster of three small flares
that form along the interface shown in panel b). Dashed lines (at $x = 10.35\,\mathrm{Mm}$ and $y = 5.6\,\mathrm{Mm}$) show the position 
of the 2D slices in Figure 2.}
\end{figure*}
We solve the 3D time-dependent, resistive MHD equations in Cartesian geometry, using the
Bifrost code \citep[]{gudiksen11}. External resistivity is computed using a 
hyper-diffusion operator \citep[]{gudiksen11}, which focuses dissipation in regions with 
steep field gradients. 
Our experiment is performed in a model that includes the upper
convection zone (with a depth of $z = -2.5\,\mathrm{Mm}$ below the photosphere), 
the photosphere/chromosphere ($z = 0-2.5\,\mathrm{Mm}$ and temperature 
$T=5\times10^{3}\,- O(10^{5}\,\mathrm{K}$, 
the transition region where the temperature increases rapidly with height, and 
the ($O(10^{6})\,\mathrm{K}$) corona that starts at $z \approx 4\,\mathrm{Mm}$ above the photosphere. 

The numerical domain
is $24 \times 24 \times 17\,\mathrm{Mm}$ in the transverse ($x$), longitudinal 
($y$) and vertical ($z$) directions respectively and it is resolved by 
$504 \times 504 \times 496$ grid points. 
In height, the resolution is $\approx 20\,\mathrm{km}$ from the 
photosphere to the lower corona increasing to $\approx 100\,\mathrm{km}$ in the upper corona and 
lower convection zone. In $x$ and $y$, the resolution is uniform ($\approx 48\,\mathrm{km}$). 
In the present experiments, convection is 
driven by optically thick radiative transfer from the photosphere. 
Radiative losses in the lower chromosphere include scattering and are assumed 
optically thin at greater heights \citep[]{carlsson12}. Field aligned thermal conduction is included.
The model is initialized with a uniform oblique magnetic field ($B < 0.1\,\mathrm{G}$) that fills 
the corona, making an inclination angle of 45 degrees with respect to the $z$-axis. 
Firstly, the experiment runs until a relatively steady state equilibrium is achieved. 
Then, to model flux emergence, a uniform magnetic flux sheet with $B_{y}=3360\,\mathrm{G}$ is inserted 
into the lower boundary, within the domain $[x,y]=[0-24,3-16]\,\mathrm{Mm}$, for a time period of $105\,\mathrm{min}$.

\begin{figure}[t]
\centering
\includegraphics[width=0.99\linewidth]{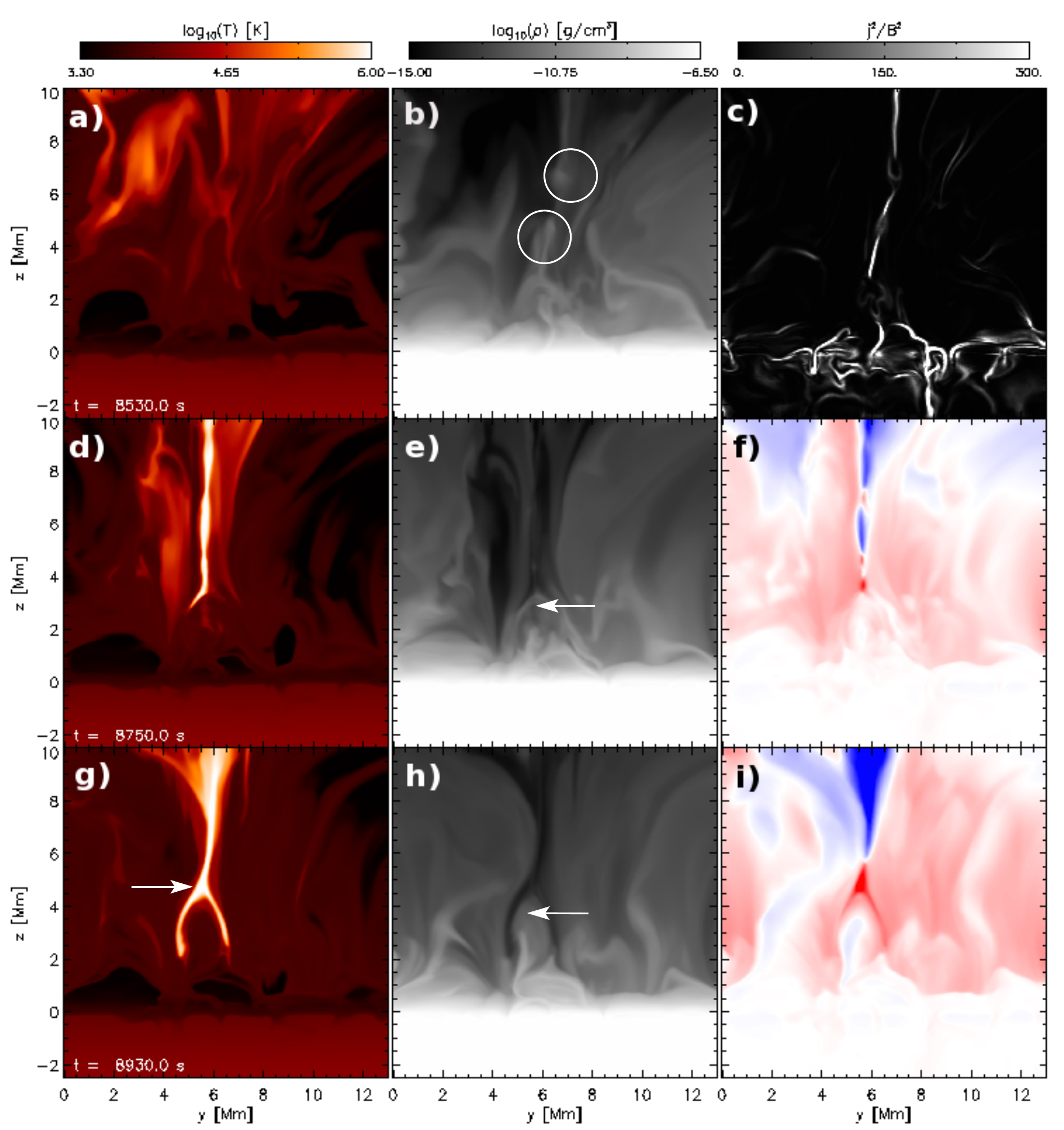}\caption{
Evolution across the interface at $x = 10.35\,\mathrm{Mm}$. 
Times are $t = 8530\,\mathrm{s}$, $t = 8750\,\mathrm{s}$, and $ t= 8930\,\mathrm{s}$ (top to bottom). Panels f) and i) show 
$v_\mathrm{z}$ in the range $[-100,100]\,\mathrm{km}\,\mathrm{s}^{-1}$ (red/blue). Two plasmoids are outlined in panel b). 
Arrows (panels e, h) show the post-flare loop and the approximate site of the flare ($z\approx 4.7\,\mathrm{Mm}$, panel g).}
\end{figure}
\begin{figure}[t]
\centering
\includegraphics[width=0.99\linewidth]{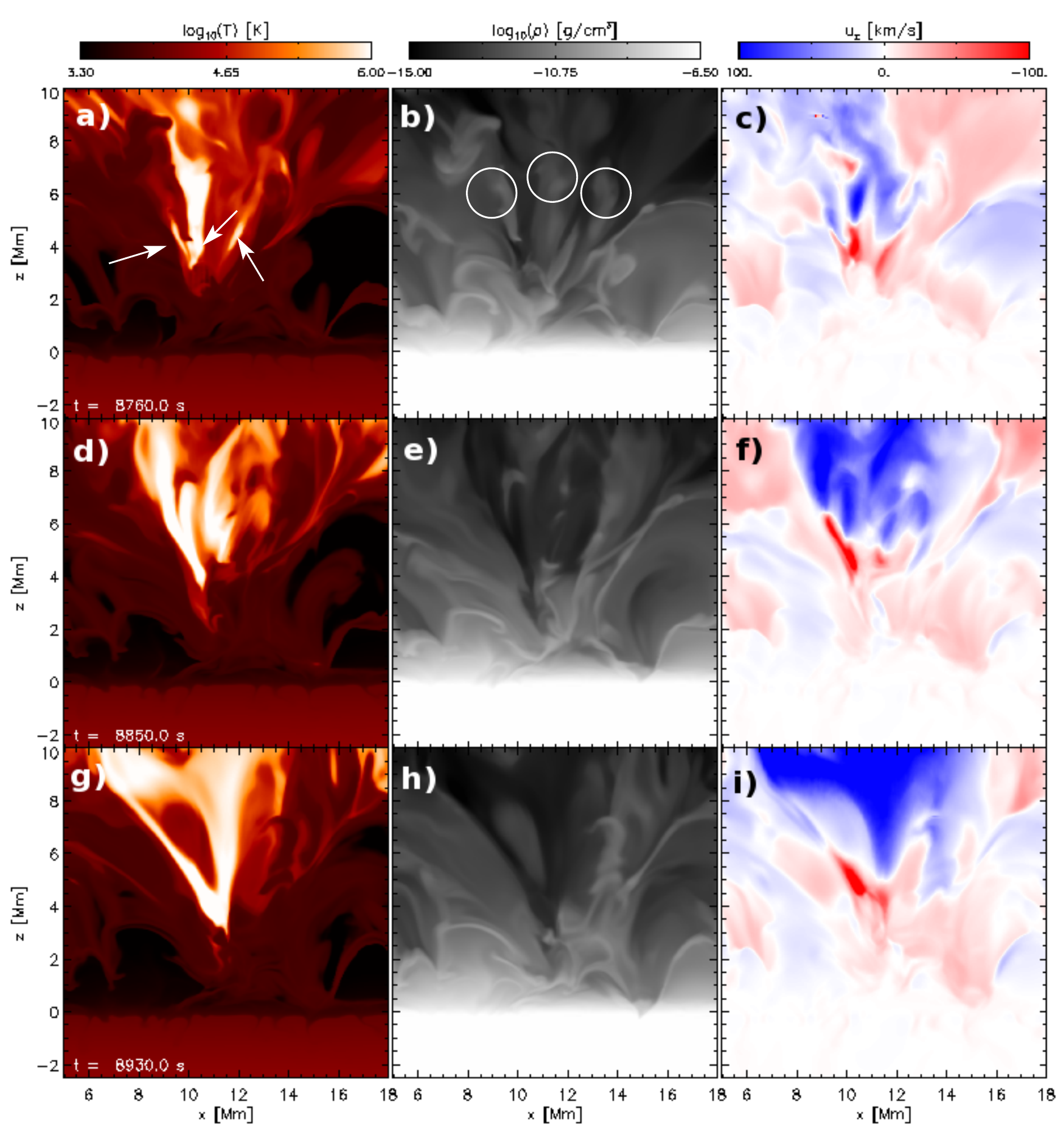}\caption{
Evolution along the interface at $y = 5.6\,\mathrm{Mm}$.
Times are: $t = 8760\,\mathrm{s}$, $t = 8850\,\mathrm{s}$, and $ t= 8930\,\mathrm{s}$ (top to bottom).
Segments of three plasmoids are outlined in panel b). The same plasmoids are visualized in the 3D space in Figure 4a.
Underneath the plasmoids, there are individual small flare sites (shown by arrows in panel a).}
\end{figure}

\section{Results and discussion}

The embedded magnetic flux sheet is distorted by convective upflows/downflows developing loops, which eventually emerge 
to the photosphere as magnetic bipoles with mixed polarity. The emergence of the magnetic field above the photosphere 
is due to the onset of the magnetic buoyancy instability, as has been shown in previous
experiments of magnetic flux emergence \citep[e.g.][]{arc04, sykora08}.
Figure 1a shows that the emergence and the expansion of the bipoles above the 
photosphere produces a multi-scale network of magnetic loops (e.g. A ($\approx 1-2\,\mathrm{Mm}$) and C ($\approx 7-8\,\mathrm{Mm}$)). 
Reconnection between the emerging loops and the oblique pre-existing field opens the way for the plasma ejecta, which occur
during the evolution of the system, to move towards the outer solar atmosphere.

Several of the magnetic loops are inter-connected via the same fieldlines in a ``sea-serpent'' manner.
Profound interaction occurs between magnetic bipoles that expand and press against each other (Figure 1b).
Their oppositely directed fieldlines come closer together and strong current layers are built up at their interface. 
The field across the interface is strongly sheared. The resistivity is locally enhanced in the current layers 
and efficient reconnection of the sheared field leads to dissipation of energy and triggering of small flares (Figure 1c).

The onset of a flaring event is illustrated in Figure 2, which shows the plasma evolution across the interface current (visualized in Figure 1b). 
For $t\ge 8530\,\mathrm{s}$, the interacting magnetic loops have expanded into the corona. Figure 2a shows the cool adiabatic 
expansion of the two interacting loops (e.g. $y\approx 2\,\mathrm{Mm}$, $y\approx 10\,\mathrm{Mm}$ and $z\approx 4\,\mathrm{Mm}$).
At the interface, the current develops into a very long and thin layer (Figure 2c), which becomes subject to the 
resistive tearing instability leading to the formation of cool and dense (Figure 2a,b) magnetic ``islands'' (i.e. plasmoids, 
\citep[e.g.][]{tsuneta97,shibata01,arc06}) 
with X-type reconnection points in between. Eventually (Figure 2e,h), the plasmoids move out of the current layer and flux from the emerging
loops comes into the interface and reconnects. This facilitates the eruption of plasmoids, the emission of successive 
bi-directional flows (jets) (Figure 2f,i) and the disclosure of a small-scale but intense brightening (i.e. flaring event, Figure 2g) 
where the plasma is heated to X-ray temperatures due to Joule dissipation via reconnection. 
After the eruption of the plasmoids we measure the ratio of the inflow towards the diffusion region to the Alfven speed 
around the reconnection site (see e.g. \citet[]{arc07} for a similar study), which is an indication of the reconnection rate. The average value 
of this ratio during the dynamical evolution of the interface is $\approx 0.1$, which indicates that the reconnection is fast. 
Typical values of plasma $\beta$
, around the reconnection regime at the interface, lie in the range 0.01-0.1.

The upward product of reconnection is a jet with a velocity of $\approx 200\,\mathrm{km}\,\mathrm{s}^{-1}$ (Figure 2i) 
and temperature up to $\approx 2.5\,\mathrm{MK}$. 
This emission can be observed as a (soft) X-ray jet.
The reconnected fieldlines that are released downwards form a small post-flare loop (Figure 2e,h), which is heated at the top ($\approx 2 \times 10^{6}\,\mathrm{K}$) 
by the collision of the downward reconnection jet with the local plasma. We find that reconnection generates slow-mode shocks that 
are attached to the ends of the diffusion region (e.g. at $z\approx 5.5\,\mathrm{Mm}$, Figure 2i). 
They exist close (and along) to the edges of the two downflows (shown in Figure 2i),
which adopt an overal inverse V-shape configuration. The slow shocks is a characteristic feature of the Petschek-type reconnection. 
Most of the conversion of the magnetic energy into heat and bulk kinetic energy (via reconnection) occurs at these shocks. 
We also find a fast termination shock, which is the result of the collision between the
downward reconnection outflow and the post-reconnection loops. In Figure 2i, the termination shock is found at 
$z\approx 4.5\,\mathrm{Mm}$. Slow and fast shocks are formed repeatedly during the dynamical evolution of the interface.
The energy released by reconnection is transported along 
the reconnected fieldlines via thermal conduction, causing flaring of the loop (Figure 2g) and heating of the 
transition region/chromospheric plasma at its footpoints. As magnetic reconnection proceeds in the corona, newly reconnected 
fieldlines successively pile up on the post-flare loop, which consequently adopts a cusp-like shape (Figure 2h,i, see also Figure 3d). 
The lifetime of the event (from flaring to cooling by radiation and conduction) is 
$\leq 100\,\mathrm{s}$ and the corresponding energy release is $10^{25}-10^{26}\,\mathrm{ergs}$.

A similar to the above process has been invoked by observations \citep[]{masuda94,tsuneta97} and theoretical 
models \citep[e.g.][]{sweet69,kopp76,shibata95} to explain large (length $\approx 100\,\mathrm{Mm}$) 
standard (e.g. two-ribbon) flares associated with the eruption of filaments or large plasmoids. 
Here, we have shown that this mechanism occurs also on much smaller scales ($\approx 1-2\,\mathrm{Mm}$) and that it 
might account for the onset of nano/micro-flares. 
Moreover, the present model shows that not only one but a cluster of small flares occur along the 3D current layer. 
This is illustrated in Figures 3 and 4.  At one time, several plasmoids can exist in the same interface. 
They are identified by e.g. a local enhancement in density (encircled in Figure 3b). 
Their ejection leads to patchy reconnection, which in turn gives onset to spatially intermittent heating (Figure 3c,f) and initiates
the individual small flares (e.g. Figure 3a, arrows). 

Due to the tearing instability and the dynamic evolution (including coalescence) of the 
plasmoids, the current layer undergoes fragmentation. It breaks up into hot ``filaments'' of intense current  (shown later in Figure 4c), which in fact are 
smaller dissipation sites emitting distinct bi-directional flows (Figure 3c,f). The above results could explain naturally the 
intermittent heating in the active solar corona \citep[e.g.][]{machado88}. On larger scales, a similar mechanism might operate to produce 
the observed multiple downflows above flare loops \citep[e.g.][]{asai04,mckenzie09}. 
These plasma properties cannot be explained by the traditional flare models, which rely on the existence of single (``monolithic'') current sheets. 

The further evolution of the system along the interface reveals an interesting result. 
For $t\ge 8850\,\mathrm{s}$, the individual bursts start to become indistinguishable from one another 
(Figure 3d,g). Since the distance between them is very short, the onset and evolution of one flare affects (even 
stimulates) the other by reconfiguring the nearby magnetic field topology. As a result, we are witnessing a gradual blending of the small flares 
and the development of another (larger in energy release) flare. 
Therefore, the explosive flaring at $t=8930\,\mathrm{s}$ (Figure 3g) is not an all-time individual brightening but it is the composite 
effect of the adjacent small flares. Indeed, the onset of the flare at $t=8930\,\mathrm{s}$ is spatially and temporally associated with the 
{\it joint} eruption of the neighbouring plasmoids. The eruption is thrusted by a V-shaped reconnection jet (Figure 3i).

\begin{figure}[t]
\centering
\includegraphics[width=0.98\linewidth]{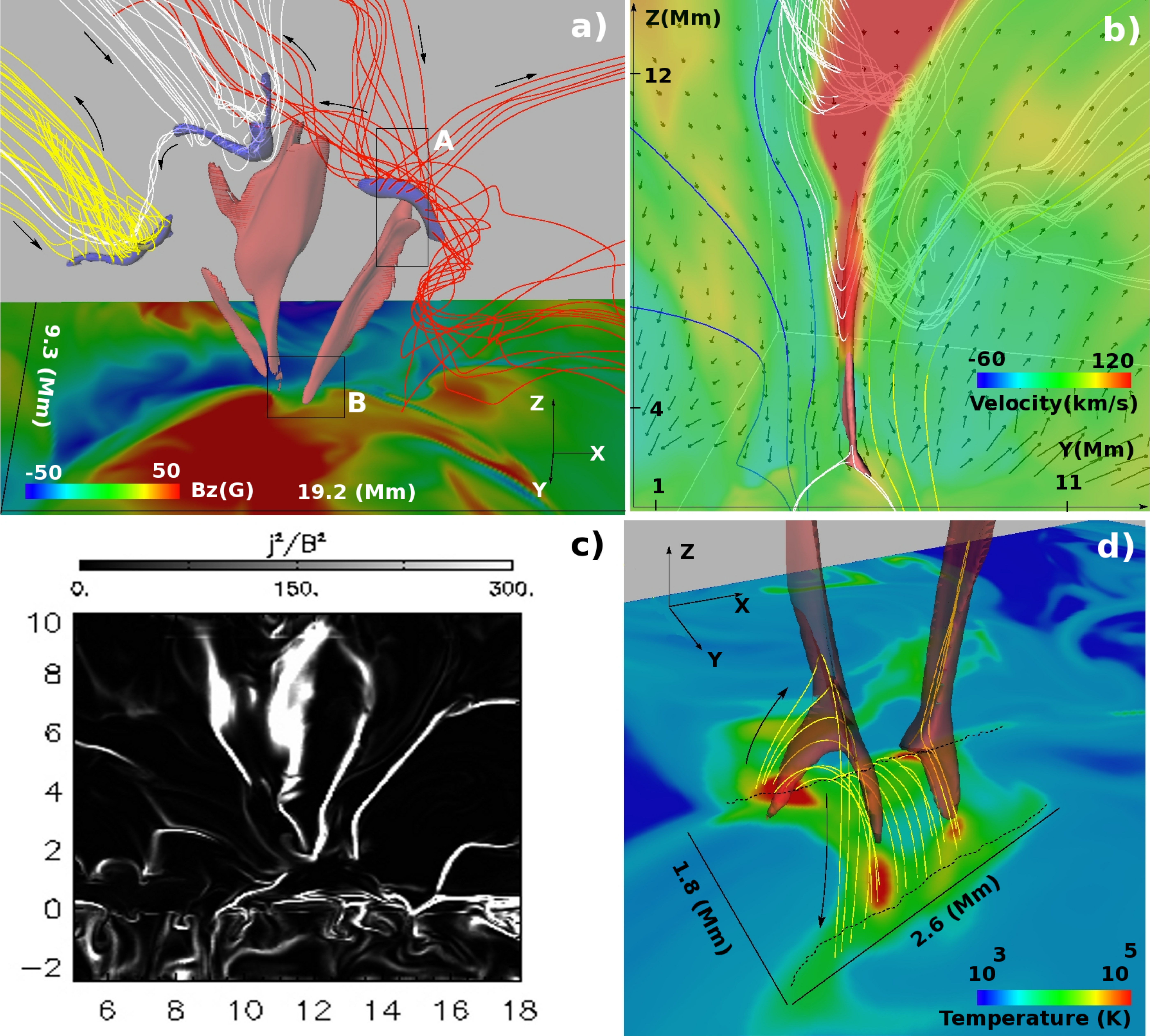}\caption{
{\bf a)} The intermittent heating 
(isosurfaces with $T\ge 5.5 \times 10^{5}\,\mathrm{K}$; red) at the interface and the erupting plasmoids 
(isosurfaces of $\rho \approx 10^{-12}\,\mathrm{g}\,\mathrm{cm}^{-3}$; blue) 
at $t=8800\,\mathrm{s}$. The (black) arrows show the direction of the field along the fieldlines. 
The horizontal slice is the Bz distribution at $z=3.14\,\mathrm{Mm}$, demonstrating that the heating occurs between opposite polarity fields. 
{\bf b)} Close-up of the region around the plasmoid, which is outlined by the frame A in panel a). 
The B-field vector (arrows) shows the highly sheared field across the interface. 
Fieldlines (yellow, blue belong to the interacting loops) reconnect and a fast jet is emitted underneath 
the erupting plasmoid (twisted fieldlines, white) and above the flare loops (lower white fieldlines). 
{\bf c)} 2D vertical cut at the interface ($y=5.6\,\mathrm{Mm}$) showing $J^{2}/B^{2}$. Time is $t=8850\,\mathrm{s}$.
{\bf d)} Close-up of the temperature isosurfaces (framed at inset B, panel a). 
The cusp-shaped reconnected fieldlines (in yellow) show the topology of the post-flare loops. 
The dashed lines indicate the length of the small two-ribbon flare.} 
\end{figure}

Figure 4a shows the eruption of plasmoids and the intermittent heating at the interface in the 3D space. 
The (blue) isosurfaces illustrate the dense core of the  plasmoids. Underneath, there are the hot filamentary structures of the 
fragmented current layer. Small flares are produced at the lower part of each filament (see later in Figure 4d). The magnetic fieldlines show the highly 
twisted nature of the field in and around the plasmoids. The individual plasmoid fieldlines reconnect rapidly where they intersect and, thus, they 
become stochastic as they pass through successive plasmoids: fieldlines (e.g. white and red) go through the close vicinity 
of (at least) two plasmoids in succession. This implies that the eruption of one plasmoid can affect or even initiate the eruption of 
the others. Accordingly, each flare can change the fieldline topology so that other sites start to reconnect and another flare 
is set off nearby. This can lead to ``sympathetic'' flaring activity and, ultimately, to profound heating and plasma expulsion as we discussed 
earlier. 

The eruption of plasmoids evolves into helical jets. This is partially illustrated
in Figure 4b. The upward fast reconnection jet (red) consists of the twisted
fieldlines (white) of the plasmoid on its upper part (at $z\approx 12\,\mathrm{Mm}$).
During the emission, the erupted plasmoids come into contact with the ``open'' ambient
(non-twisted) magnetic field and reconnect. This leads to the formation of helical jets and
an untwisting motion along the jets (for similar studies see e.g. \citet[]{shimojo07,arc13}).
Figure 4b also shows that the hot reconnection plasma underneath the plasmoid is ejected upward 
by the tension of the reconnected (V-shaped, white) fieldlines at the interface. During the evolution, 
similar reconnection events occur at various atmospheric heights (from the chromosphere up to the corona) and, thus, we detect the 
emission of several EUV and X-ray jets from the interface. Many of the hot reconnection jets, which are produced at the 
interface, have velocities comparable to the local Alfven speed.

Figure 4c displays $J^{2}/B^{2}$ at $y=5.6\,\mathrm{Mm}$ (the same vertical cut as in Figure 3). It is shown that the interface current 
is not uniform and coherent but it is fragmented. It consists of thin, individual, filamentary structures of strong current where efficient reconnection occurs. 
A comparison with Figure 4a shows that there is a good spatial relationship between the sites of profound heating and the places with strong current. 
This implies that the intermittent heating is due to the fragmentation of the interface current layer.

Figure 4d displays the lower part of the hot filaments (inverse Y-shaped temperature isosurfaces, $T\ge 5 \times 10^{5}\,\mathrm{K}$). 
Reconnected fieldlines form an arcade of loops. The overal arcade is not the result of a single eruptive flare, as has been previously 
thought, but it consists of distinct, cusp-like, small eruptive flares, which show up as ``miniature'' version of bigger flares 
\citep[e.g.][]{masuda94,tsuneta96,tsuneta97}. Two bright ribbons form on either side of the
neutral line, joining the successive footpoints of the post-flare loops. Thus, a cluster of small eruptive flares can form a tiny 
two-ribbon flare. 

\begin{figure}[t]
\centering
\includegraphics[width=0.98\linewidth]{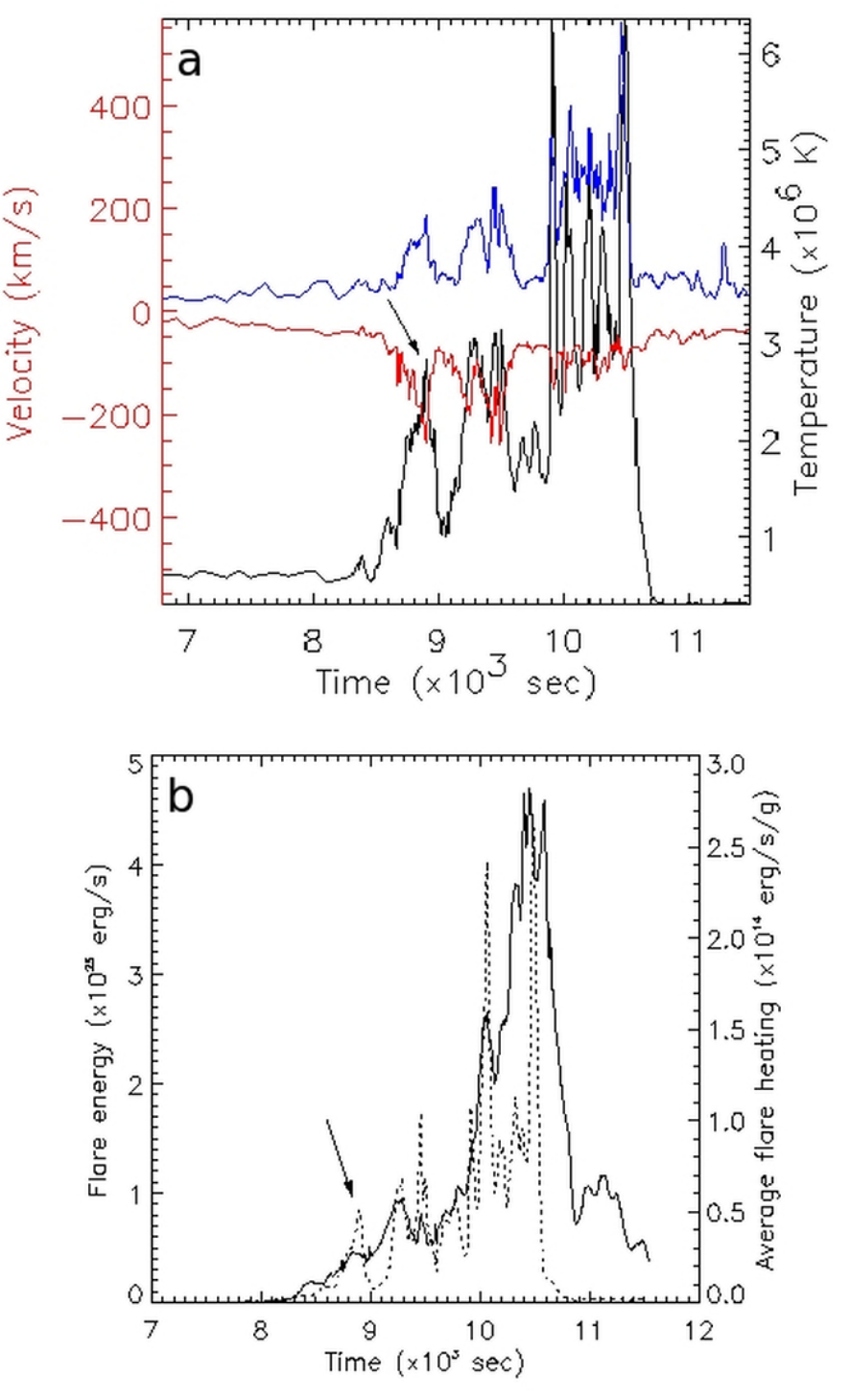}\caption{
{\bf a)} Temporal evolution of maximum temperature and maximum (blue)/minimum (red) vertical velocity 
at the volume interface that 
surrounds the small flares: $x=[8.6,12.9]\,\mathrm{Mm}$, $y=[2.4,7.1]\,\mathrm{Mm}$, $z=[2,10]\,\mathrm{Mm}$. 
{\bf b)} The same for the total flare energy per second (solid line) and the average flare energy per unit mass (dashed line). 
The arrow indicates the explosive reconnection event at $t=8900\,\mathrm{s}$.}
\end{figure}

The powering of the flares at the interface occurs repetitively over a $40\,\mathrm{min}$ period: we find more than 20 flares  
at $t\approx 8400-10800\,\mathrm{s}$.
The heat and energy input (Figure 5a, b) to the chromosphere/corona is highly intermittent, with profound fluctuations in temperature as the 
plasma is suddenly heated by small flares and subsequently cools down. During the above period, there is a remarkable temporal (and spatial) 
correlation between heating, energy release and the onset of fast ($200-400\,\mathrm{km}\,\mathrm{s}^{-1}$) jets. 
This is a direct evidence of reconnection-driven plasma acceleration in small eruptive flares. Similar evolution is found in 
other interfaces within the numerical domain. 
The produced flares appear at random intervals, with an average lifetime of $\approx 30\,\mathrm{s}-3\,\mathrm{min}$. They occur at various atmospheric heights 
(chromosphere-corona) and they are capable of heating the plasma to $\approx 1-6\,\mathrm{MK}$. Some of the larger spikes in Figure 5b (e.g. 
$t\approx 10050\,\mathrm{s}$, $t\approx 10500\,\mathrm{s}$ etc., solid line) represent individual energy emissions of $O(10^{27})\,\mathrm{ergs}$. 
These events might account for microflares. 
However, many of the events with noticeable 
total energy release (e.g. around $t\approx 8900\,\mathrm{s},9300\,\mathrm{s},10400\,\mathrm{s}$) is the result of the superposition of small flares, each involving 
$10^{25}-10^{26}\,\mathrm{ergs}$, which is the nano/micro-flare energy regime.

For the small flares that occur in the corona, we estimate that the average energy flux in the area of integration 
(i.e. $4.3\times 4.7\,\mathrm{Mm}$, as in Figure 5) is at least $~O(10^{6})\,\mathrm{erg}\,\mathrm{s}^{-1}\,\mathrm{cm}^{-2}$. 
This estimate, together with the high occurence rate of the flares in the same area, indicates that nano/micro-flares 
can provide a non-negligible contribution of heating in emerging flux regions 
and in the active X-ray corona.
Moreover, we estimate that the fast upward propagation of plasma, which originates mainly from the flare regimes, carries a vast amount of Poynting flux into 
the corona (in the range $1-60\,\mathrm{kW}\,\mathrm{m}^{-2}$ over the whole domain, during $t\approx 8400-10800\,\mathrm{s}$), part of which could 
contribute to the mass loading and driving of the solar wind. 
We anticipate that the mechanism presented in our numerical experiments may constitute a generic process, which powers eruptive flaring activity 
of magnetic fields in astrophysical and laboratory plasmas.

\acknowledgments
This research was supported by the Research Council of Norway through the grant 
``Solar Atmospheric Modelling'' and through grants of computing time from the Programme for Supercomputing, by the European 
Research Council under the European Union's Seventh Framework Programme (FP7/2007-2013) / ERC Grant agreement nr. 291058 and 
by computing project s1061 from the High End Computing Division of NASA. The authors are grateful for in-depth discussions 
during the ISSI workshops: ``Magnetic flux emergence in the solar atmosphere'' (Bern, February 2011) and 
``Understanding solar jets'' (Bern, March 2013).

\bibliographystyle{apj}

\clearpage

\end{document}